\documentclass[fleqn,usenatbib]{mnras}

\usepackage{newtxtext,newtxmath}

\usepackage[T1]{fontenc}
\usepackage{ae,aecompl}

\usepackage{graphicx}	
\usepackage{amsmath}	
\usepackage{amssymb}	
\usepackage{gensymb}
\usepackage{hyperref}

\title[White Dwarf Pollution]{White Dwarf Pollution by Asteroids from Secular Resonances}

\author[Smallwood et al.]{
Jeremy L. Smallwood,$^1$\thanks{E-mail: smallj2@unlv.nevada.edu}
Rebecca G. Martin,$^1$
Mario Livio$^1$
and Stephen H. Lubow$^2$
\\
$^1$Department of Physics and Astronomy, University of Nevada, Las Vegas, 4505 South Maryland Parkway, Las Vegas, NV 89154, USA\\
$^2$Space Telescope Science Institute, Baltimore, MD 21218, USA
}

\date{Accepted XXX. Received YYY; in original form ZZZ}

\pubyear{2017}

\begin{document}
\label{firstpage}
\pagerange{\pageref{firstpage}--\pageref{lastpage}}
\maketitle

\begin{abstract}
In the past few decades, observations have revealed signatures of metals polluting the atmospheres  of white dwarfs. The diffusion timescale for metals to sink from the atmosphere of a white dwarf is of the order of days for a hydrogen-dominated atmosphere.  Thus, there must be a continuous supply of metal-rich material  accreting onto these white dwarfs. 
We investigate the role of secular resonances that excite the eccentricity of asteroids allowing them to reach star-grazing orbits leading them to tidal disruption and the formation of a debris disc.
Changes in the planetary system during the evolution of the star lead to a change in the location of secular resonances.  In our Solar System, the engulfment of the Earth will cause the $\nu_6$ resonance to shift outwards which will force previously stable asteroids to undergo secular resonant perturbations.  With analytic models and $N$--body simulations we show that secular resonances driven by two outer companions can provide a source of continuous pollution.  Secular resonances are a viable mechanism for the pollution of white dwarfs in a variety of exoplanetary system architectures. 
\end{abstract}



\section{Introduction}
About $30\%$ to $50\%$ of all white dwarfs have metal-polluted atmospheres identified by metallic absorption lines from spectroscopic measurements and a total of about $1000$ white dwarfs are known to be polluted \cite[e.g.][]{Cottrell1980,Koester1982,Lacombe1983,Zeidler1986,Koester1997,Zuckerman2003,Klein2010,Vennes2010,Zuckerman2010,Farihi2012,Farihi2014,Melis2017}. These findings appear at first glance to be puzzling since white dwarf atmospheres stratify chemical elements through gravitational settling \citep{Koester2006,Koester2009}. Once the white dwarf has cooled below $25,000\, \rm K$, metals are no longer supported by radiative forces and rapidly sink and diffuse within  the white dwarf's atmosphere due to the intense gravity environment \citep{Fontaine1979,Vauclair1979,Koester2009}.
It is noteworthy that white dwarfs are observed in a temperature range from $5,000$ K to $25,000$ K \citep{Fontaine2001}, which coincides with cooling ages of $2 \, \rm Gyr$ to $20\, \rm Myr$, respectively \citep{Koester2014}.
The stratification or diffusion timescales for metals are of the order of days to weeks for DA (hydrogen atmosphere) white dwarfs and $10^4-10^6$ yrs for DB (helium atmosphere) white dwarfs \citep{Koester2006}.  This implies that the diffusion timescale of metals is many orders of magnitude shorter than the white dwarf cooling time \citep{Paquette1986}. Accretion discs and pollution are observed at $30\, \rm Myr$ to $600\, \rm Myr$ cooling ages \citep{Farihi2016}. Thus, these polluted white dwarfs need to be continuously accreting metal-rich material in order for the metal absorption lines to be observable.

Several theoretical models have been explored to explain the metal pollution. Accretion of metallic material originating from the interstellar medium has been conclusively ruled out by \cite{Farihi2010b} \cite[see also][]{Aannestad1993,Jura2006,Kilic2007,Barstow2014}. 
The currently favored pollution mechanism suggests that metal-rich planetary material is tidally disrupted (due to close encounters with the star) into a debris disc and then subsequently accreted onto the white dwarf \citep{Gansicke2006,Kilic2006,vonHippel2007,Farihi2009,Jura2009,Farihi2010,Melis2010,Brown2017,Bonsor2017,xu2018}. For a recent review on the dynamics of white dwarf pollution, see \cite{Farihi2016}.  The sources of pollution that have been proposed include asteroids \citep{Jura2003,Jura2006,Jura2009,Debes2012,Veras2013,Wyatt2014}, comets \citep{Caiazzo2017}, moons via planet-planet scattering \citep{Payne2016,Payne2017} and perturbations of planetary material due to eccentric planets \citep{Frewen2014}. Perturbations may also be caused by Kozai-Lidov instabilities in stellar binaries \citep{Hamers2016,Petrovich2017}. \cite{Veras2017} recently formulated the critical separation of binaries in order for the stellar companion to produce pollution of white dwarf atmospheres by Roche lobe overflow or by stellar winds. Their findings suggested that the critical separation is only a few astronomical units (au), 
which means that this mechanism is inefficient for wide binary systems.


The chain of events is thought to be as follows. As a main-sequence star evolves, the star undergoes significant mass loss during the red-giant branch (RGB) phase \citep{Reimers1977,McDonald2015} and during the asymptotic giant branch (AGB) phase \citep{Rosenfield2014,Rosenfield2016}. This mass loss is a result of radiation pressure pushing on the loosely bound outer layers of the red giant. The mass outflow, or stellar wind, leads to mass-loss rates of up to $10^{-4}\, \rm M_{\odot} yr^{-1}$ \cite[e.g.][]{veras2011} that depletes the star of a large fraction of its initial mass. During the expansion of the outer envelop, close-in planets are engulfed \citep{Siess1999,Villaver2007,Villaver2009,Mustill2012,Adams2013,Villaver2014}, leading (among other things) to dynamical changes within the system. As the star undergoes significant mass loss, due to conservation of angular momentum, the orbits of all surviving bodies expand outward \citep{Duncan1998}. For planets and asteroids that are at orbital radii well below  a few hundred au of the star, the timescale for mass loss is much longer than the orbital periods of the planets and so the orbits of the planets and asteroids expand adiabatically \citep{Veras2013}.

To take our Solar system as an example, many asteroids are located in the asteroid belt between Mars and Jupiter. Over time, some asteroids undergo resonant gravitational perturbations from the two largest planets, Jupiter and Saturn, causing the asteroids to become scattered from the asteroid belt \citep{Morbidelli1995,Gladman1997,Morbidelli1998,Bottke2000,PetitMorbidelli2001,Ito2006, Bro2008, Minton2010, Chrenko2015,Granvik2017}. In a mean-motion resonance, the ratio of the orbital periods of two objects is an integer ratio \cite[e.g.][]{Armitage2013}. Secular resonances arise when the apsidal or nodal precession rates of two objects orbiting a central object are close \citep{Froeschle1986,Yoshikawa1987}. Apsidal resonances are more important in the context of white dwarf pollution since apsidal resonances excite eccentricities, which can lead to tidal disruptions, whereas nodal resonances excite inclinations. The most important apsidal secular resonance in our solar system is the $\nu_6$ resonance \cite[e.g.][]{Bottke2000,Ito2006, Minton2011,Haghighipour2016,smallwood2017}, that occurs between the apsidal precession of the asteroids and Saturn. The outer edge of the $\nu_{6}$ resonance sets the inner boundary to our asteroid belt at approximately $2.1\,\rm AU$. The region where Jupiter's mean-motion resonances overlap is what determines the outer edge of the asteroid belt at about $3.3\,\rm AU$ to $3.5\, \rm AU$.  
Each mean-motion resonance has a width, in semi-major axis, over which it operates \citep{Dermott1983}. The locations and widths of secular resonances are harder to constrain due to their strong dependences on the semi-major axis, eccentricity, and inclination \citep{Knezevic1991}. We use a first-order analytical approximation in section~\ref{sec:analytic} to determine the locations and eccentricity excitation regions of these resonances. Asteroids close to the resonance location undergo perturbations causing their eccentricities to increase, eventually leading the asteroids to either be ejected from the solar system or collide with the central star. Regions in which the resonance widths overlap are known as chaotic regions \citep{Murray1997, Murray1999}, and there almost all of the asteroids are cleared out.

\cite{Debes2012} investigated white dwarf pollution by asteroids that originate from the 2:1 mean-motion resonance with Jupiter, which is presently located at $3.276\, \rm au$ \citep{Nesvorny1998,PetitMorbidelli2001}. In their model, as the star loses mass through its stellar evolution, the libration width of the 2:1 resonance is slightly widened, forcing previously stable asteroids to eventually become accreted onto the white dwarf. We propose that secular resonances may provide an additional mechanism and source of pollution. 
If the planetary system undergoes major changes during the stellar evolution (for instance, if the inner planets are engulfed by the star), the secular frequencies of the whole planetary system are affected, resulting in a displacement of the location of secular resonances. On the other hand, the locations of mean-motion resonances with respect to the remaining planets remains almost unchanged. \cite{Ward1981} discuses a similar process whereby secular resonances are affected by the mass loss of the solar nebula rather than the central star. The changes in the dynamics of these resonances may have aided in the accretion of planetesimals by growing terrestrial planets.


In the present work we investigate the evolution of an exoplanetary system that contains a white dwarf that harbors a planetary system and an asteroid belt. The giant outer planets and the asteroid belt are sufficiently far from the white dwarf so that they survive the stellar evolution through the RGB/AGB phases. We consider systems with two giant planets, like the solar system, and systems with one giant planet and a binary stellar companion. We explore how different system architectures are able to pollute the atmospheres of white dwarfs. In Section 2 we describe the analytical and numerical models that we use to calculate the location and dynamics of secular resonances that will occur within our Solar System as it evolves. In Section 3 we  analyze white dwarf pollution for various architectures of exoplanetary systems. Finally, we draw our conclusions in Section 4.  

\section{Solar System}
In this Section we first consider how our solar system will evolve once our Sun is on its way to becoming a white dwarf. We assume that the terrestrial planets, up to the orbital radius of the Earth, will become engulfed by the Sun \citep{Rasio1996,Schroder2008}, while the orbits of the giant planets and the asteroid belt will expand adiabatically.  We model the evolution of the  $\nu_6$ resonance in the solar system first analytically and then numerically with $N$--body simulations of the asteroid belt with the remaining planetary system. Even though Mercury and Venus will also be engulfed, we focus on the effects of the Earth engulfment. Because the Earth is more massive and located closer to the $\nu_6$ resonance it affects it more strongly.

\subsection{Analytic Model}
\label{sec:analytic} 

Here we examine the resonance location and the eccentricity excitation region for the $\nu_6$ resonance in the solar system both before and after the Sun becomes a white dwarf. The present-day values for the orbital elements are used for Earth, Jupiter, and Saturn. The $\nu_6$ secular resonance properties are mostly affected by both Saturn and Jupiter \citep{Bottke2000,Ito2006}. Jupiter increases the free precession frequency of the asteroids so that they fall into a resonance with an eigenfrequency that is dominated by Saturn. In this paper, the location of the $\nu_6$ resonance is estimated by calculating the location of the intersection of a test particle's free precession rate with the eigenfrequency dominated by Saturn. The analytical model we use is linear in eccentricity and inclination and it gives the secular perturbations at first order to the orbital perturbation.

\subsubsection{Eigenfrequencies}
\label{sec:eig}
We consider a planetary system with a total of $N$ planets orbiting a central object with mass $m_*$. Each planet has a semi--major axis $a_j$, mass $m_j$ and orbital frequency $n_j = \sqrt{Gm_* / a_j^3}$, where $j=1,...,N$. The eigenfrequencies are found by calculating the eigenvalues of the $N\times N$ matrix $A_{jk}$ associated with a generalized form of the secular perturbation theory
\begin{equation}
A_{jk} = -\frac{1}{4} \frac{m_k}{m_* + m_j}n_j \alpha_{jk}\bar{\alpha}_{jk}b^{(2)}_{3/2}(\alpha_{jk}) 
\label{A1}
\end{equation}
for $j\ne k$ and otherwise
\begin{equation}
A_{jj} = \frac{n_j}{4} \sum_{k=1,k\neq j}^{N} \frac{m_k}{m_* + m_j}\alpha_{jk}\bar{\alpha} _{jk}b^{(1)}_{3/2}(\alpha_{jk})
\label{A2}
\end{equation}
\citep{MurrayBook2000,Minton2011,malhotra2012}, where the Laplace coefficient $b_{s}^{(j)}(\alpha)$ is given by
\begin{equation}
\frac{1}{2}b_{s}^{(j)}(\alpha) = \frac{1}{2\pi}\int_0^{2\pi}\frac{\cos (j\psi)\, d\psi}{(1-2\alpha \cos \psi + \alpha^2)^s}
\end{equation}
and  $\alpha_{jk}$ and  $\bar{\alpha}_{jk}$ are defined as
\begin{equation}
  \alpha_{jk}=\begin{cases}
    a_k/a_j, & \text{if $a_j > a_k$ \quad (internal perturber)},\\
    a_j/a_k, & \text{if $a_j < a_k$ \quad (external perturber)},
  \end{cases}
\end{equation}
and
\begin{equation}
  \bar{\alpha}_{jk}=\begin{cases}
    1, & \text{if $a_j > a_k$ \quad (internal perturber)},\\
    a_j/a_k, & \text{if $a_j < a_k$ \quad (external perturber)}.
  \end{cases}
\end{equation}

We find that  the $g_6$ eigenfrequency has a value of $22.13^{\prime \prime} \rm yr^{-1}$ (includes only Jupiter and Saturn) and a value of $22.16^{\prime \prime} \rm yr^{-1}$ (includes all the planets in the Solar System), which is lower by roughly $20\%$ from the more accurate value of $27.77^{\prime \prime} \rm yr^{-1}$ given by \cite{Brouwer1950} (see \cite{Laskar1988} for further comparisons).

\subsubsection{Asteroid free precession rates}

We calculate the free precession rate of test particles in the potential of the planetary system. In this linear theory in eccentricity and inclination, it is only a function of the secular semi-major axis $a$.  The free precession rate is given by
\begin{equation}
g_0 = \frac{n}{4}\sum_{j=1}^N\frac{m_j}{m_*}\alpha_j \bar{\alpha}_j b_{3/2}^{(1)}(\alpha_j)
\label{A}
\end{equation}
\citep[e.g.][]{MurrayBook2000}, where $n$ is the orbital frequency of the test particle. The variables $\alpha_j$ and $\bar{\alpha}_j$ are defined as
\begin{equation}
  \alpha_{j}=\begin{cases}
    a_j/a, & \text{if $a_j < a$},\\
    a/a_j, & \text{if $a_j > a$},
  \end{cases}
\end{equation}
\begin{equation}
  \bar{\alpha}_{j}=\begin{cases}
    1, & \text{if $a_j < a$},\\
    a/a_j, & \text{if $a_j > a$}.
  \end{cases}
\end{equation}
The mean precession frequency, $g_0$, corresponds to the diagonal term of the Laplace-Lagrange matrix including the asteroid \citep{Milani1990,morbidelli1991}. In this work, we will first consider the case $N=3$ (Earth and Jupiter and Saturn), and after that take $N=2$ (Jupiter and Saturn) since the inner planets are engulfed during the RGB/AGB phases. The outer giants Neptune and Uranus do not significantly affect the dynamics of the asteroid belt \citep[e.g.][]{Izidoro2016}. Saturn also does not noticeably affect the {free} precession rate of the asteroids in the asteroid belt -- that rate is dominated by Jupiter.  



\begin{table*}
\caption{Parameters used to calculated the  maximum forced eccentricity of a test particle during main-sequence and post-main-sequence, as shown in Fig.~\ref{width}. The columns beginning from left to right are as follows: parameter description, symbol, followed by the parameter value for the two evolutionary stages, main-sequence and post-main sequence.}
\begin{tabular}{llll} 
	\hline
	Parameter & Symbol & Main-sequence value & Post-main-sequence value \\
    \hline
    Star Mass& $M_{\rm Star}/M_{\odot}$  & $1$ &$0.5$ \\
    Earth Mass& $M_{\rm E}/M_{\odot}$  & $3.04 \times 10^{-6}$ & --- \\
    Jupiter Mass& $M_{\rm J}/M_{\odot}$  &  $0.00095786$& $0.00095786$ \\
    Saturn Mass& $M_{\rm S}/M_{\odot}$ & $0.000285837$ & $0.000285837$ \\
    Earth semi-major axis& $a_{\rm E}/{\rm au}$ & $1$ & ---\\
   Jupiter semi-major axis& $a_{\rm J}/{\rm au}$ & $5.20$ & $10.40$ \\
    Saturn semi-major axis& $a_{\rm S}/{\rm au}$ & $9.55$ & $19.10$ \\
    Earth Eccentricity& $e_{\rm E}$ & $0.0167$ & --- \\
    Jupiter Eccentricity& $e_{\rm J}$ & $0.0475$ & $0.0475$ \\
    Saturn Eccentricity& $e_{\rm S}$ & $0.0575$ & $0.0575$ \\
    Earth longitude of perihelion& $\bar{\omega}_{\rm E}/\degree$ & $102.94719$  & ---\\
    Jupiter longitude of perihelion& $\bar{\omega}_{\rm J}/\degree$ & $13.983865$  & $13.983865$ \\
    Saturn longitude of perihelion& $\bar{\omega}_{\rm S}/\degree$ & $88.719425$ & $88.719425$ \\
    Location of $\nu_6$ resonance & $a_{\nu_6}/{\rm au}$ & $1.81$  & $3.68$\\
    \hline
	\end{tabular}
    \label{table1}
\end{table*}

\subsubsection{Resonance location}

\begin{figure}
\includegraphics[width=8.4cm]{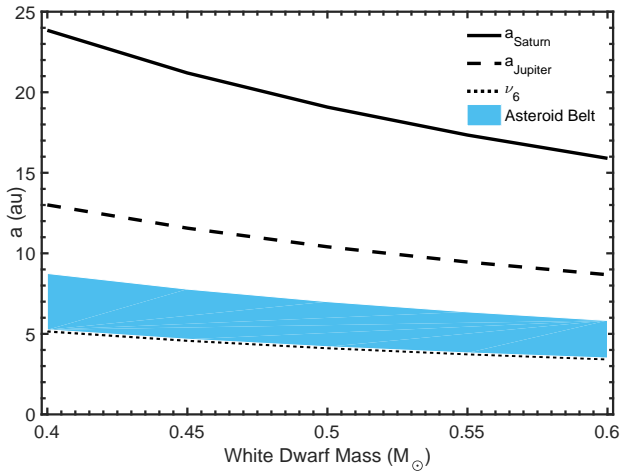}
\caption{The semi-major axes of Saturn, Jupiter, and the $\nu_6$ secular resonance as a function of white dwarf mass. The location of the $\nu_6$ resonance is found by calculating the location where the asteroid's free precession rate is equal to Saturn's dominant proper mode. The semi-major axes of Jupiter and Saturn depend on the adiabatic expansion which is proportional to the ratio of the initial stellar mass to the white dwarf mass (see equation~\ref{ad}). The location of the adiabatically expanded asteroid belt as a function of white dwarf mass is shown by the blue-shaded region. }
\label{sum}
\end{figure}

The semi-major axes of the planets  are assumed to undergo adiabatic expansion based on the ratio of the initial stellar mass to  the white dwarf mass, 
\begin{equation} 
a_{\rm final} =  a_{\rm initial}\left(\frac{m_*}{m_{\rm wd}}\right).
\label{ad}
\end{equation}
Figure~\ref{sum} shows the location of Jupiter, Saturn, and the $\nu_6$ secular resonance with the proper mode of Saturn as a function of white dwarf mass.  We consider white dwarf masses in the range  $0.4\, \rm M_{\odot}$ to $0.6\, \rm M_{\odot}$ as expected for the Sun \citep{Liebert2005,Falcon2010,Tremblay2016}. For our standard model we choose a mass of  $0.5\, \rm M_{\odot}$ \citep[e.g.][]{Sackmann1993,Schroder2008}.  The location of the adiabatically expanded asteroid belt as a function of white dwarf mass is shown by the blue-shaded region. We use the observed inner and outer boundaries of the asteroid belt to produce this region, that is the range $[2.1;3.5]\, \rm au$ in semi-major axis. The observed inner boundary, created by the $\nu_6$ resonance, is consistent with our analytical model which places it at about $2\, \rm au$.

\subsubsection{Maximum Forced Eccentricity}

\begin{figure}
\includegraphics[width=8.2cm]{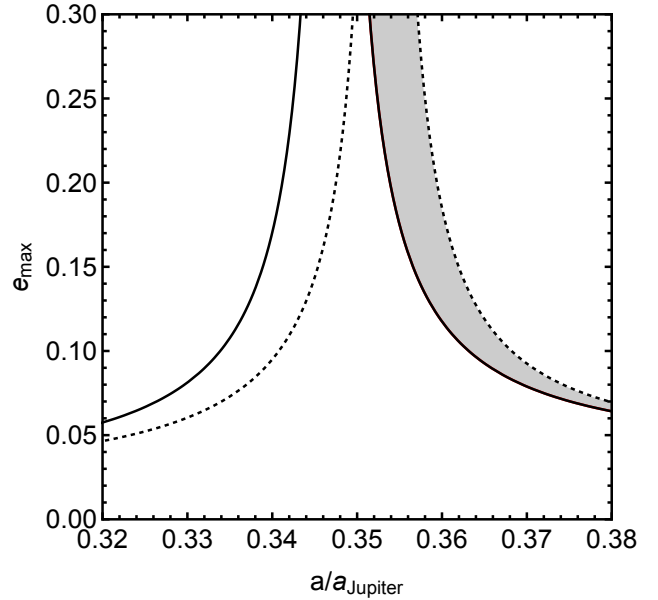}
\caption{The maximum forced eccentricity as a function of the semi-major axis of a test particle, showing the eccentricity excitation region of the $\nu_6$ secular resonance during the main-sequence stage (solid-black lines) versus post-main-sequence stage (dotted-black lines). The eccentricity excitation region during the main-sequence was calculated with the planets, Earth, Jupiter, and Saturn. As the star evolves during the post-main-sequence stage, we assume the Earth is engulfed. This engulfment shifts the $\nu_6$ resonance outwards relative to the asteroids, allowing previously stable asteroids to undergo stronger secular oscillations of eccentricity.
The shaded region represents the region of previously stable asteroids that undergo increased eccentricity growth due to secular resonant perturbations. The analytic theory is not accurate for such high values of the eccentricities, but we show it as an indication.}
\label{width}
\end{figure}

\begin{figure*} \centering
\includegraphics[width=8.6cm]{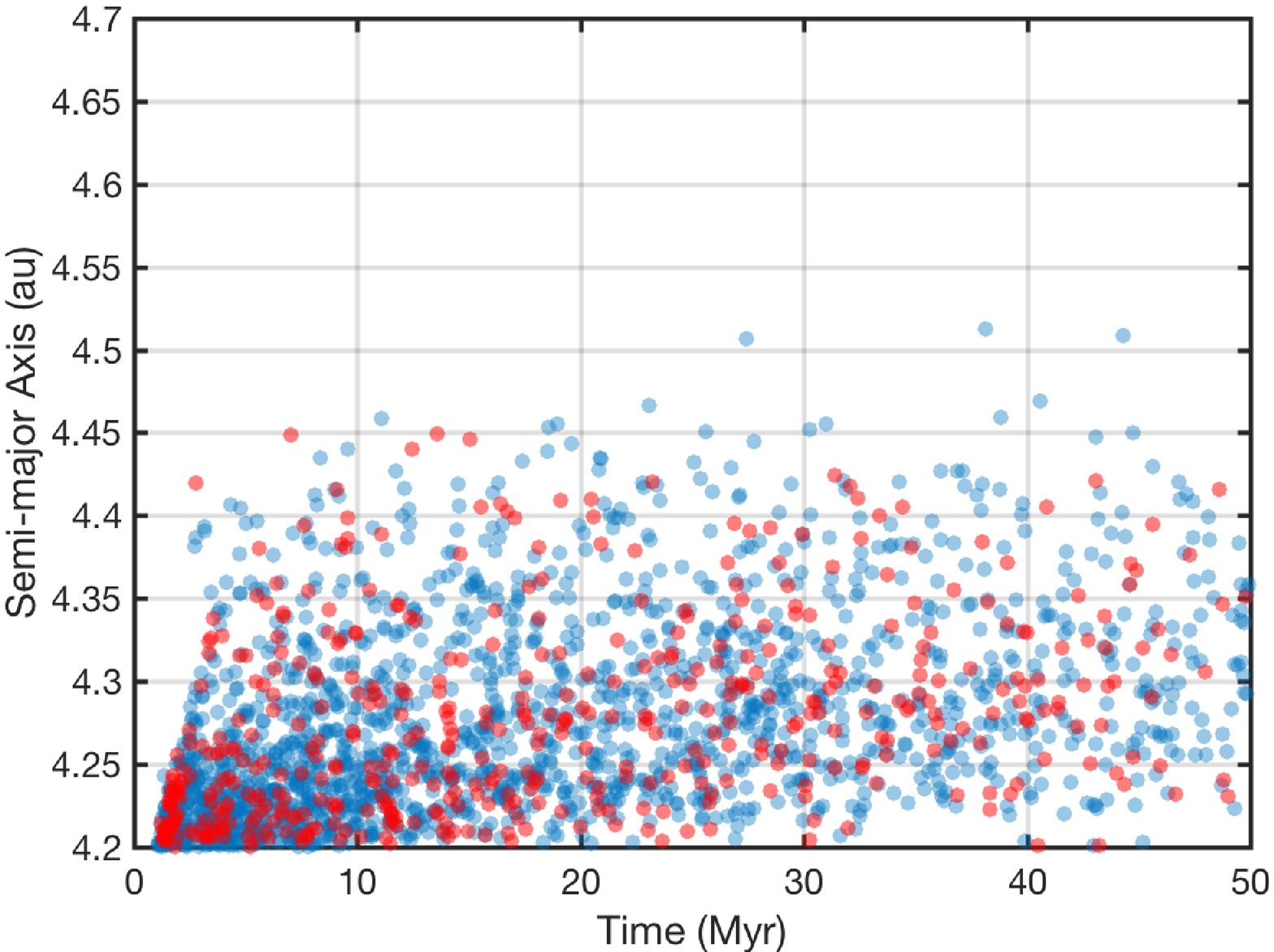}
\includegraphics[width=8.6cm]{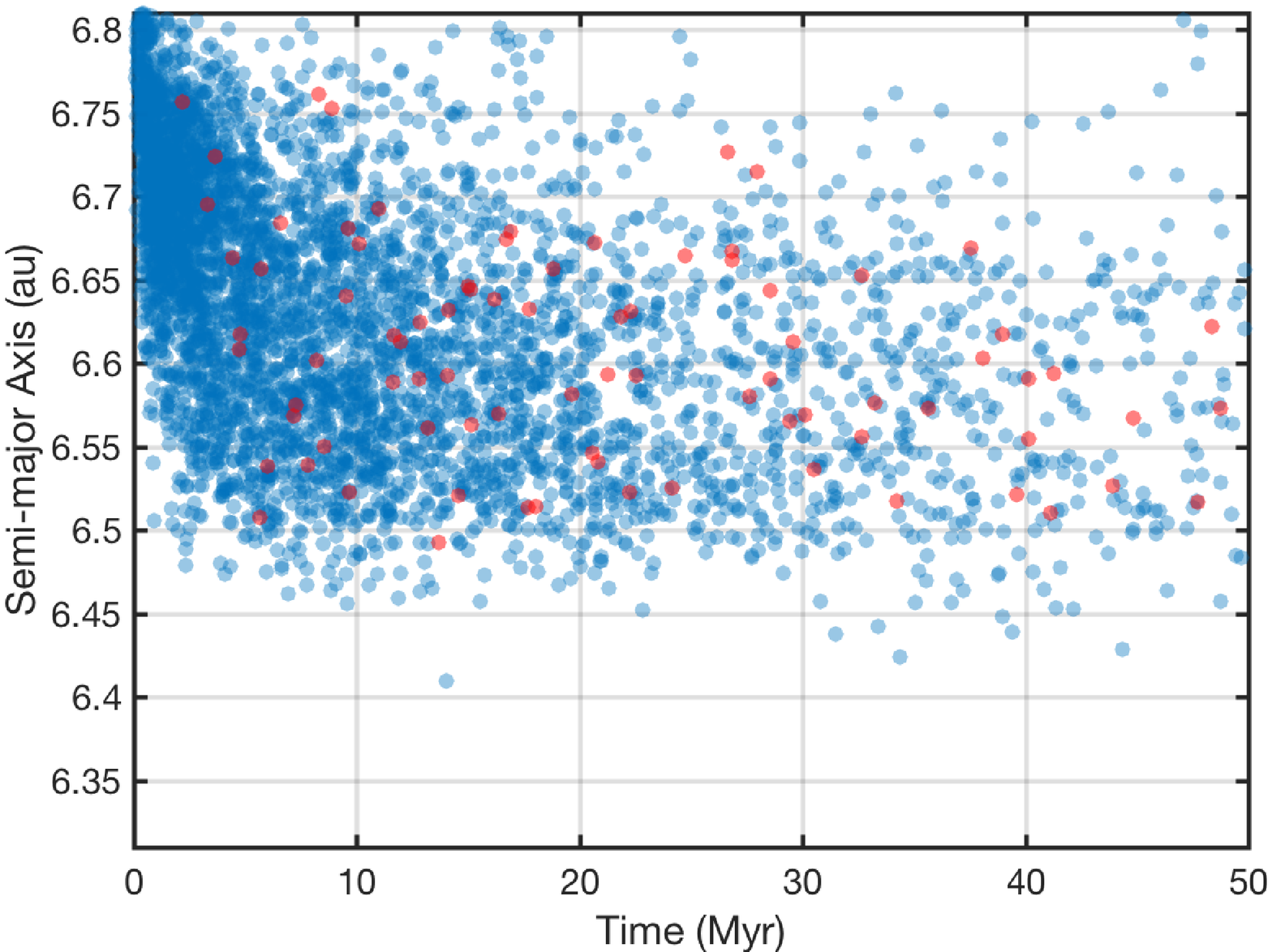}
\caption{4-body simulations (Sun, Jupiter, Saturn and asteroid) of the evolution of a set of asteroids near the $\nu_6$ secular resonance (left panel) and near the 2:1 mean-motion resonance (right panel) around a $0.5\,\rm M_{\odot}$ white dwarf. The potential outcomes for an asteroid includes ejection (blue dots), tidal disruption (red dots) or remains within the distribution. The initial semi-major axis of the test particles is given with the time of either ejection or tidal disruption.  Each simulation is comprised of 20,000 test particles initially distributed uniformly over a width of $0.5\, \rm au$. Test particles experience the same frequency ratios during adiabatic expansion. Since the radii of mean-motion resonances varies inversely with the mass of the central star, the majority of asteroids located within the 2:1 resonance would have been depleted by the time the star evolved to become a white dwarf. In our simulation, we assume there is no depletion in order to compare the number of tidal disruption events due to the mean motion resonance with the number due to the $\nu_6$ resonance. The inner boundary of the $\nu_6$ simulations was produced by knowing that the observed inner boundary of our asteroid belt is located at $2.1 \, \rm au$. 
}
\label{outcome}
\end{figure*}


In order to calculate the forced eccentricity excitation of a test particle near the $\nu_6$ secular resonance, we follow the procedure in Chapter 7 of \cite{MurrayBook2000}. We begin with the eccentricity part of the disturbing function, $V_{\rm ecc}^{\rm sec}$, from the secular theory for $N$ planets including a test particle with mean motion, $n$, eccentricity, $e$ and longitude of the perihelion, $\bar{\omega}$, given by
\begin{equation}
V_{\rm ecc}^{\rm sec} = na^2 \bigg[\frac{1}{2}g_0 e^2  + \sum_{j=1}^N A_j e e_j \cos(\bar{\omega} - \bar{\omega}_j)\bigg],
\end{equation}
where $g_0$ is the test particle free precession rate given in equation~(\ref{A}) and
\begin{equation}
A_j = -n \frac{1}{4}\frac{m_j}{m_{*}}\alpha_j \bar{\alpha}_j b_{3/2}^{( 2)}(\alpha_j).
\end{equation}
The forced eccentricity is given by
\begin{equation}
e_{\rm forced} = \sqrt[]{h_0^2(t) + k_0^2(t)},
\end{equation}
where
\begin{equation}
h_0(t) = -\sum_{i=1}^N\frac{\nu_i}{g_0-g_i}\sin(g_i t + \beta_i)
\label{h_0}
\end{equation}
and
\begin{equation}
k_0(t) = -\sum_{i=1}^N\frac{\nu_i}{g_0-g_i}\cos(g_i t + \beta_i).
\label{k_0}
\end{equation}
The constants $\beta_i$ are determined from initial boundary conditions, $g_i$ is the dominant proper mode of the $i^{\rm th}$ planet and $\nu_i$ is given by
\begin{equation}
\nu_i = \sum_{j=1}^N A_j e_{ji},
\end{equation}
where $e_{ji}$ are the scaled eigenvector components corresponding to the eigenfrequencies found from equations~(\ref{A1}) and~(\ref{A2}), obtained from the initial conditions of the planetary system given in Table~\ref{table1}. The amplitudes $e_{ji}$ of the orbital solution for the planets also depend on the longitudes of perihelion ($\bar{\omega}$) of the planets. Thus we assume that all $\bar{\omega}$  values are taken at present-day values. Since the time dependence in equations~(\ref{h_0}) and~(\ref{k_0}) are different during the main-sequence stage versus the post-main-sequence stage, we calculate the maximum forced eccentricity
\begin{equation}
e_{\rm max} =   \sum_{i=1}^N \left\lvert \frac{\nu_i}{g_0-g_i} \right\rvert.
\end{equation}

Figure~\ref{width} shows the maximum forced eccentricity excitation of a test particle near the $\nu_6$ resonance during the main-sequence stage of stellar evolution (solid line) versus the post-main-sequence stage (dotted line).  To obtain these curves we use the parameters presented in Table~\ref{table1}. The main sequence stage includes Earth, Jupiter and Saturn at semi-major axes $1.0\, \rm  au$, $5.2\, \rm au$, and $9.5\, \rm  au$, respectively. The post-main-sequence stage only includes Jupiter and Saturn with semi-major axes of $10.4\, \rm  au$ and $19.0\, \rm  au$, respectively, assuming that the Earth is engulfed during the RGB phase of stellar evolution. The $x$-axis is normalized with respect to the semi-major axis of Jupiter in order to show the comparison. The resonance has shifted outwards into a region of the asteroid belt that would have previously contained stable asteroids. These asteroids are unstable to resonant perturbations and may be a source of pollution for white dwarfs. The amplitude of the shift is quite small compared to the crude approximation of the dynamics given by the linearized analytical model. The error within our analytical model is of some tenths of astronomical units in the location of the resonance, which is of the same order of magnitude of the shift. However, the mechanism we describe here is still qualitatively relevant. The shift could be much larger if a planet more massive than the Earth is engulfed.  We investigate this further with numerical models in the next Section.

\subsection{$N$--body Simulations}
\label{sec:numerical}
We test the analytic models of the previous section with $N$--body simulations of an asteroid belt around a white dwarf. We use the hybrid symplectic integrator in the orbital dynamics package, {\sc mercury}, to model the structure of the asteroid belt and the tidal disruption rate around a $0.5\, \rm M_{\odot}$ white dwarf. {\sc mercury} uses $N$--body integrations to calculate the orbital evolution of objects moving in the gravitational field of a large body \citep{Chambers1999}. We simulated the motion of  Jupiter, Saturn, and a distribution of asteroids orbiting a white dwarf star. The asteroids in our simulations are considered test particles which interact gravitationally with the planets and the white dwarf. We may neglect the asteroid-asteroid interactions because the timescale for such collisional interactions is much longer than the timescale for the action of perturbations by resonance effects. The timescale for resonant effects is of the order of $\sim 1 \, \rm Myr$ \citep{ItoTanikawa1999}, whereas some of the largest asteroids have collisional timescales that are of the order of the age of the solar system \citep{Dohnanyi1969}. The general outcomes of test particles near secular and mean-motion resonances include ejections, collisions with a larger body, or remains within the simulation. 

As the asteroids are scattered  from the asteroid belt due to secular resonances, the asteroids become tidally disrupted by the white dwarf if they pass within a tidal disruption radius given by 
\begin{align}
R_{{\rm tide}}  & = C_{{\rm tide}}R_{{\rm wd}}\bigg(\frac{\rho_{{\rm wd}}}{\rho_{{\rm ast}}}\bigg)^{1/3}
\notag \\
& \approx 1.3\bigg(\frac{C_{{\rm tide}}}{2}\bigg)\bigg(\frac{M_{{\rm wd}}}{0.6M_{\odot}}\bigg)^{1/3}
\bigg(\frac{\rho_{{\rm ast}}}{3 \rm g\,cm^{-3}}\bigg)^{-1/3} R_{\odot},
\end{align}
\citep{Davidsson1999,Jura2003,Bear2013}, where  $M_{{\rm wd}}$, $R_{{\rm wd}}$, $\rho_{{\rm wd}}$ are the mass, radius, and density of the white dwarf, respectively, and $\rho_{{\rm ast}}$ is the  density  of the asteroid.  $C_{{\rm tide}}$ is a numerical constant that depends on the orbital parameters of the asteroid, its rotation, and composition \citep{Davidsson1999,Jura2003}. We take $C_{\rm tide}=2$ for a solid non-synchronized asteroid \citep{Bear2013}. 
We assume the average density of the asteroids to be $3\, \rm g\ cm^{-3}$ \citep{Krasinsky2002} in order to calculate the tidal disruption radius for the $0.5\, \rm M_{\odot}$ WD to be $R_{\rm tide}=1.22\, \rm R_{\odot}$. Within our simulations, we  artificially inflate the size of the white dwarf to have a radius equal to the tidal disruption radius. When an asteroid passes within the tidal disruption radius it is considered tidally disrupted and then removed from the simulation. The destabilization of asteroids should begin as soon as the Earth is engulfed, that is, when the Sun is still a red giant. Many objects will thus be ejected or will collide with the star before it reaches its white dwarf mass and radius. The evolution timescale between the red giant and white dwarf stages is around $10^4$ years. However, the timescale for resonant perturbations is on the order of $10^6$ years, thus in this work we assume that after the Earth engulfment, the Sun instantaneously shrinks to its white dwarf state in order to show the effects on the dynamics of secular resonances. We calculate the evolution of each asteroid orbit for a duration of $50$ million years, since this is longer than the cooling age of many white dwarfs.

\subsubsection{Efficiency Comparison}
We first set up two simulations with a distribution of asteroids that is uniform in semi--major axis in order to compare the efficiency of tidal disruption events from the $\nu_6$ resonance to the 2:1 mean motion resonance.  The actual asteroid belt distribution is far from being uniform, so we analyze its actual distribution in section~\ref{tidrate}.  Each simulation has a width of $0.5\, \rm au$ in initial semi-major axis. The simulation range in semi-major axis for the $\nu_6$ resonance  simulation was taken to be  $4.2\, \rm au$ up to  $4.7\, \rm  au$.   The inner boundary of the $\nu_6$ simulation is chosen based on the adiabatic expansion (see equation~\ref{ad}) of the observed inner  boundary of the asteroid belt, which is located at $2.1\, \rm au$. Thus, we only simulate the region that we expect the resonance to operate. The inner and outer boundaries of the 2:1 simulation at  $6.31\, \rm  au$ and $6.81\, \rm au$  are chosen so as to be centered on the location of the resonance at $6.56\, \rm au$.  The value of $6.56\, \rm  au$ represents the adiabatic expansion of the average semi-major axis value of the location of the 2:1 mean-motion resonance \cite[$3.276\, \rm au$,][]{Nesvorny1998}. Note that this simulation does not include any asteroid depletion at the resonance location. 

Between these boundaries, we placed $20,000$ test particles. The orbital elements for each asteroid were chosen as follows: the semi-major axis ($a$) was sampled uniformly in the range described in the previous paragraph, the inclination ($i$) was distributed in the range $0-10\degree$, and the eccentricity ($e$) was randomly allocated from the range $0.0 - 0.1$. The remaining orbital elements, the longitude of the ascending node ($n_{as}$), the argument of perihelion ($g$), and the mean anomaly ($M_{a}$), were all randomly allocated in the range  $0-360\degree$.  The semi-major axes of Jupiter and Saturn were chosen based on adiabatic expansion (see values for a $0.5\, \rm M_{\odot}$ white dwarf in Fig.~\ref{sum}). The remaining orbital elements for the planets were taken to be equal to the present-day values, since the outer solar system is stable over long timescales \citep{Laskar1994}. 


\begin{figure} \centering
\includegraphics[width=8.4cm]{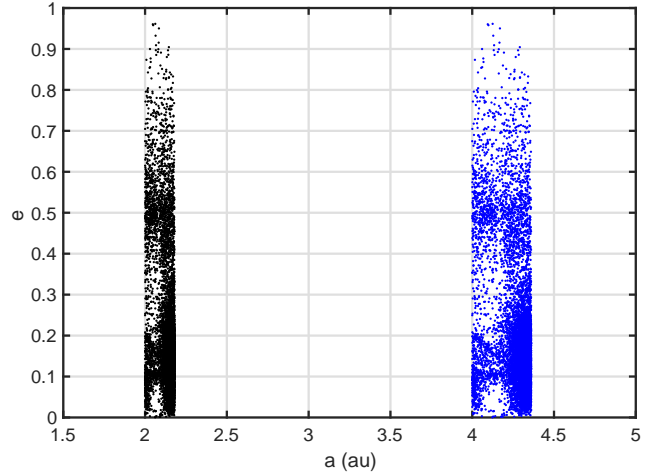}
\caption{The distribution of $10,000$ objects in the actual asteroid belt near the $\nu_6$ secular resonance presently (main-sequence, black dots) and adiabatically shifted (post-main-sequence, blue dots). Data for the present-day objects are taken from the MPC Orbit Database.}
\label{real_belt}
\end{figure}

\begin{figure} \centering
\includegraphics[width=8.2cm]{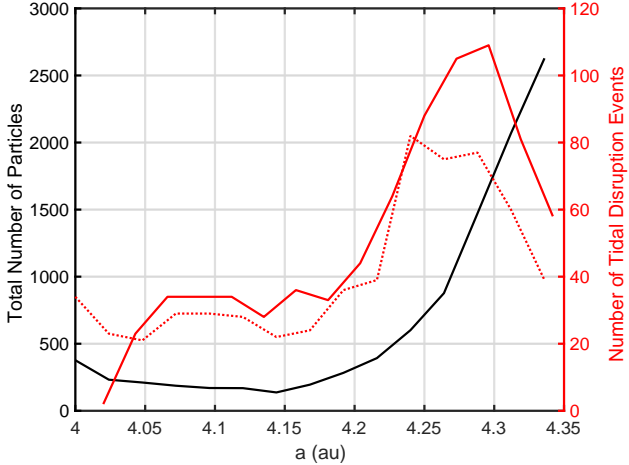}
\caption{ The black line shows the total number of particles from the initial distribution shown in Fig.~\ref{real_belt} as a function of initial semi-major axis. The red line denotes the number of tidal disruption events for the Earth engulfment simulation and the red-dotted line shows the number of tidal disruption events without Earth engulfment, both as a function of initial semi-major axis after $50\, \rm Myr$ (follows scale on the right axis). The difference in the peaks between the two red lines is contributed by the shift in the secular resonance as the Earth is engulfed.}
\label{tid_v_total_earth}
\end{figure}

\begin{figure*} \centering
\includegraphics[width=8.8cm]{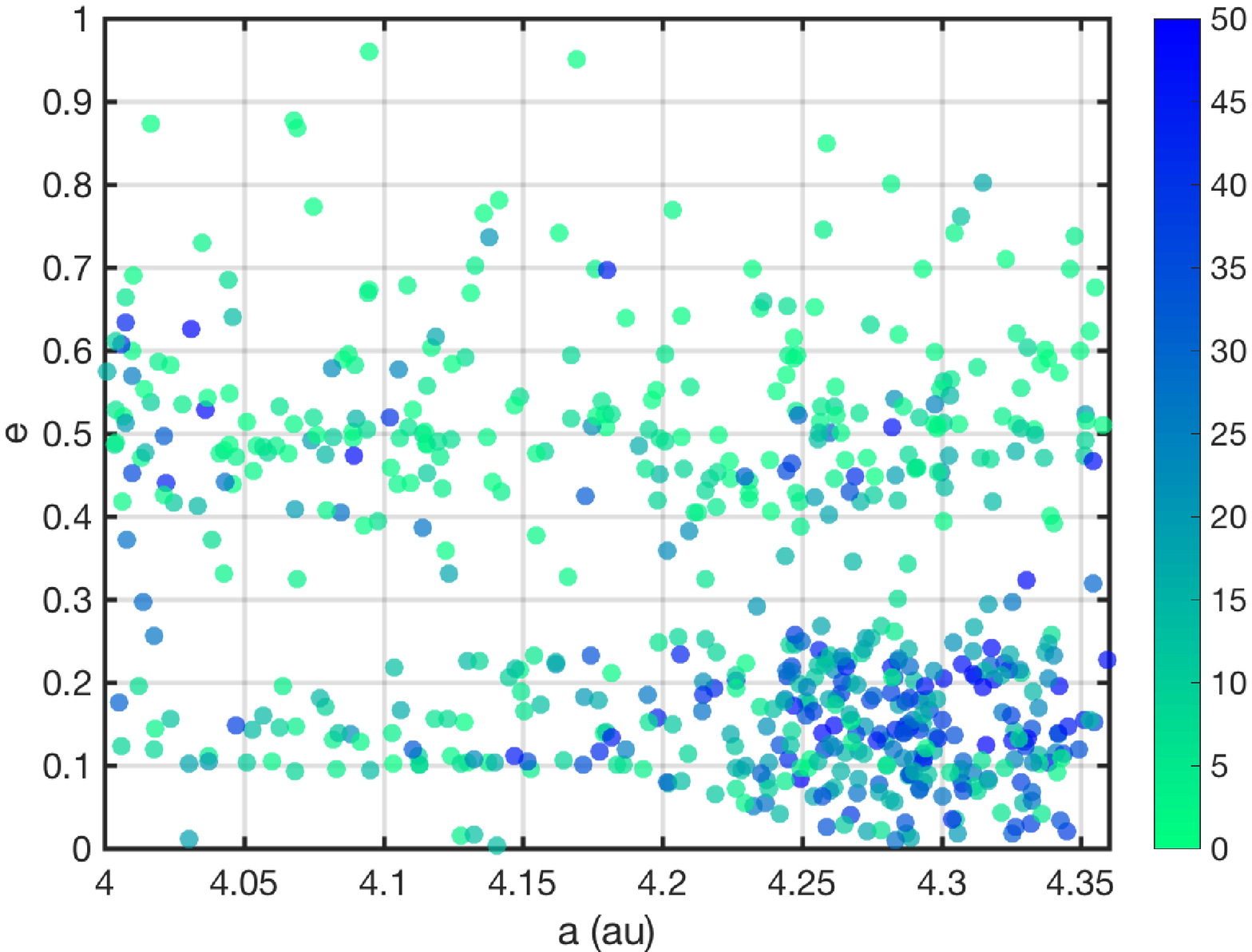}
\includegraphics[width=8.8cm]{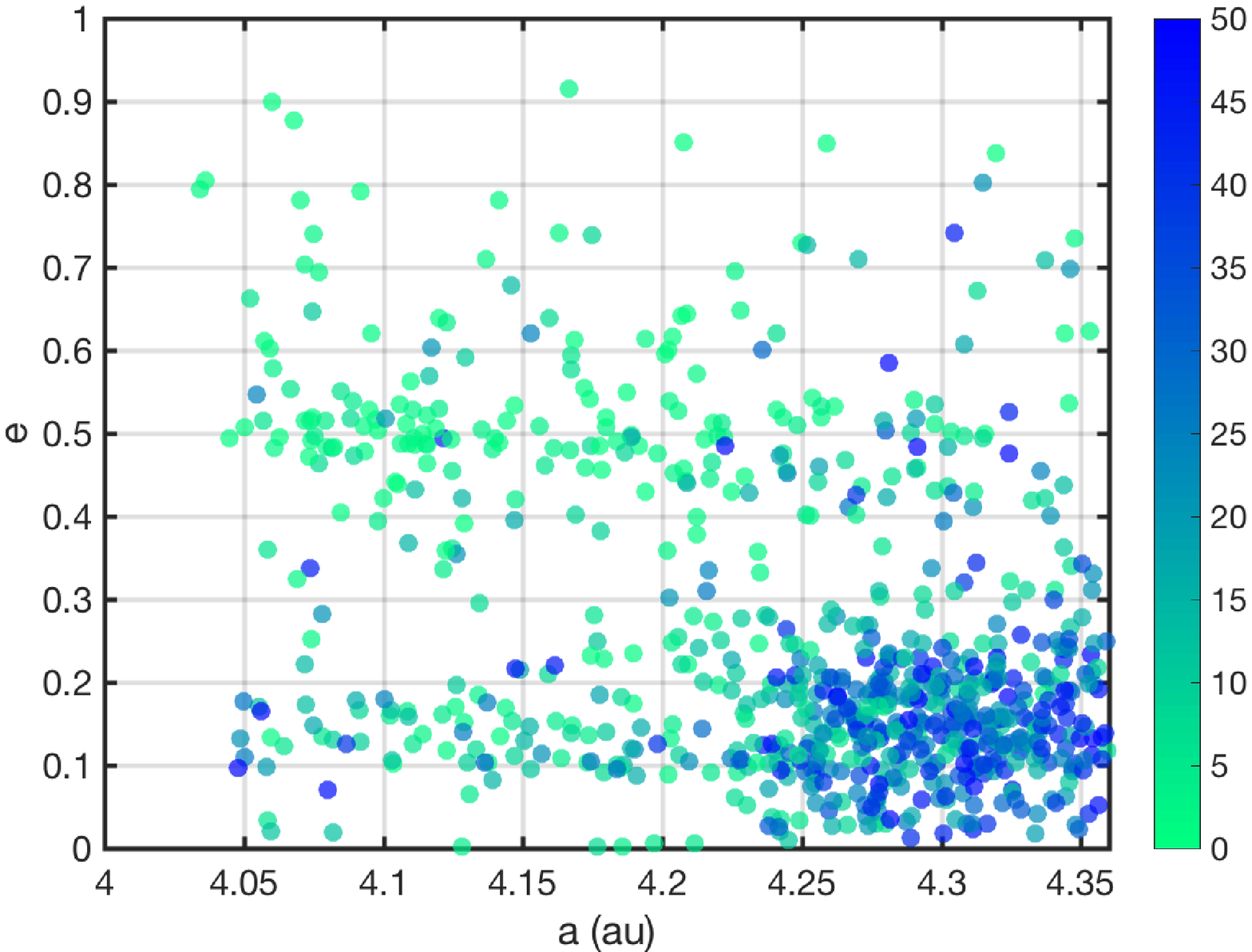}
\caption{ The initial distribution of asteroids near the present-day location of the $\nu_6$ secular resonance that were tidally disrupted during the simulation with no Earth engulfment (left panel) and with Earth engulfment (right panel). The color bar represents the time an asteroid was tidally disrupted, with  bright green at $t=0\, \rm Myr$ and dark blue at $t = 50\, \rm Myr$.}
\label{a_e_time}
\end{figure*}

\begin{figure} \centering
\includegraphics[width=8.4cm]{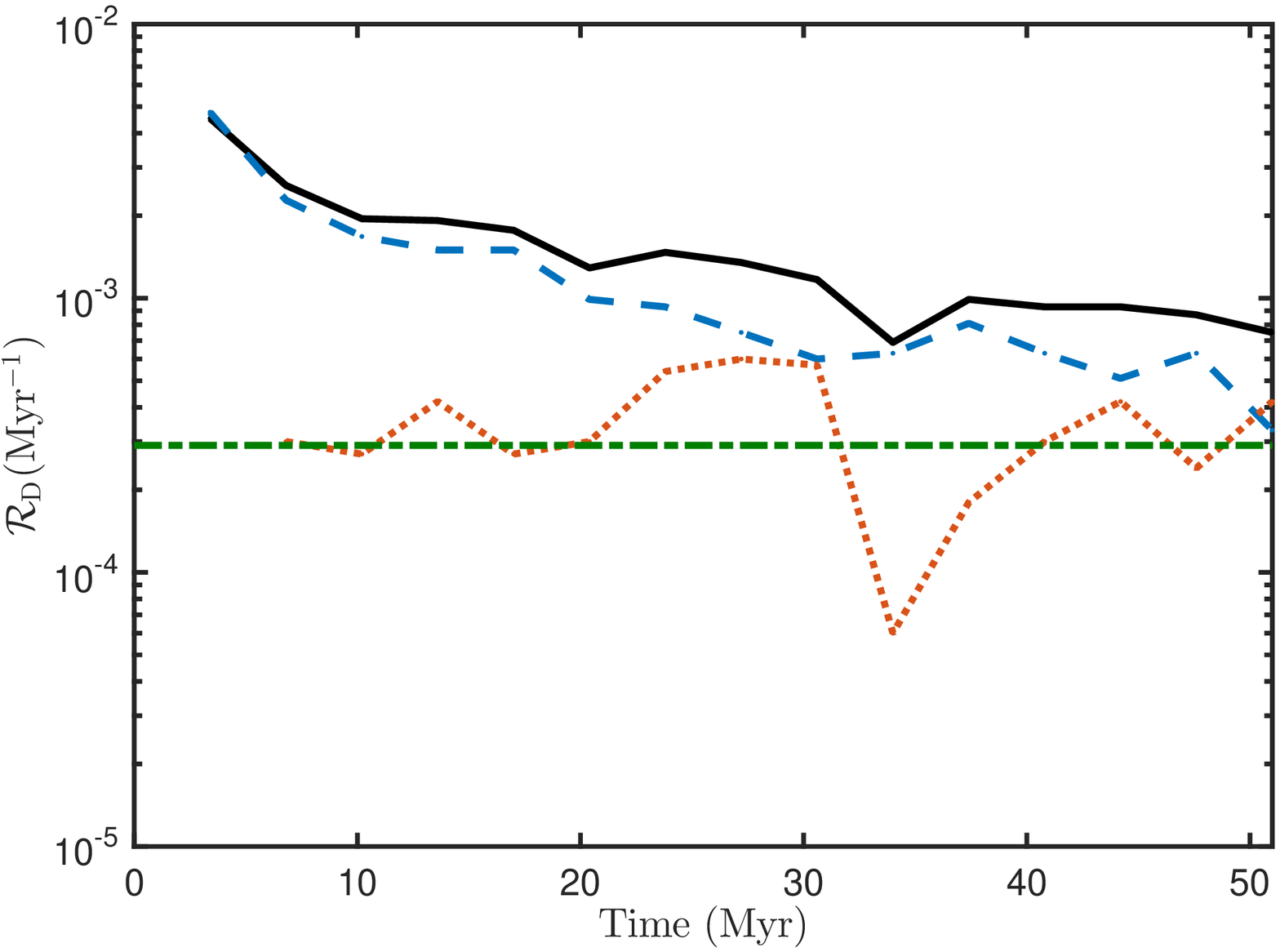}
\caption{The tidal disruption rate ($\mathcal{R_{\rm D}}$) of asteroids as a function of time for the adiabatically expanded asteroid belt near the $\nu_6$ secular resonance (see Fig.~\ref{real_belt}) around a  $0.5\,\rm M_{\odot}$ white dwarf. The black line shows the rate for the simulation with Earth engulfment and the dashed-blue line shows the rate without the Earth being engulfed. The difference between these two rates is denoted by the dotted-red line.  We also show a constant best-fitting line of the difference denoted by the dashed-dotted green line.  The number of tidal disruption events is normalized to the initial number of test particles in our simulations. }
\label{rate_hist}
\end{figure}

Figure~\ref{outcome} summarizes the results of these numerical simulations.  The outcomes for each asteroid include ejection (blue dots) and tidal disruption (red dots). Since mean-motion resonance radii vary inversely with the mass of the star during adiabatic expansion, the majority of the asteroids located within the 2:1 resonance would have been depleted by the time the star evolved to become a white dwarf. In our simulation, we have not taken account of any depletion and therefore the number of tidal disruption events is highly over--estimated. For our fiducial uniform distribution simulations the number of tidal disruption  events for the $\nu_6$ case is also over--estimated (since the asteroid belt is not uniform). However, the ratio of the number of ejections to tidal disruptions is higher for the 2:1 resonance than for the $\nu_6$ resonance. This suggests that the 2:1 mean-motion resonance is not nearly as efficient in producing tidal disruption events as is the $\nu_6$ secular resonance. 

\subsubsection{Tidal Disruption Rate from the Asteroid Belt}
\label{tidrate}
Observationally \cite{Koester2014} conducted an unbiased survey for DA white dwarf metal pollution  with cooling ages in the range of $20-200\, \rm Myr$ and temperature  $17,000 \rm K < T_{eff} < 27,000 K$. Using previous ground-based studies and adopting bulk Earth abundances for the debris discs, mass accretion rates range from a few $10^5\, \rm g\, s^{-1}$ to a few  $10^8\, \rm g\, s^{-1}$.
We now consider whether the asteroid belt in the Solar system would be able to provide an accretion rate in this range from the shift in the $\nu_6$ resonance. 

We calculate the tidal disruption rate for asteroids that originate from the location of the $\nu_6$ resonance case in the asteroid belt in the Solar system. This is an important aspect of white dwarf pollution because for this mechanism to be a major contributor there needs to be a continuous supply of asteroids. In order to disentangle the effect of the shift of the $\nu_6$ resonance by Earth engulfment from stellar evolution, we compare two simulations, one with Jupiter and Saturn and another one with Earth, Jupiter, and Saturn. To estimate the rate of tidal disruption events from the outward shift of the $\nu_6$ secular resonance, we setup  $10,000$ test particles that have the orbital properties of asteroids in the asteroid belt taken from the MPC Orbit Database\footnote{https://www.minorplanetcenter.net/iau/MPCORB.html}. We use an unbiased selection of objects, which includes near-earth asteroids (NEAs). This population of asteroids have high eccentricities and are unstable on timescales of order a million years \citep{morbidelli2002book}, thus they would not be a source of long term pollution of white dwarf atmospheres. The objects are drawn randomly for semi--major axis in the range $2 - 2.18$ au.
We evolve the test particle population that orbits a $0.5\, {\rm M_{\odot}}$ white dwarf for a time of $50\, \rm Myr$. We assume that these test particles and the planets (in both simulations) have undergone adiabatic expansion due to the stellar mass loss associated with the evolution of a white dwarf star. Figure~\ref{real_belt} shows the initial distribution of test particles used to calculate the tidal disruption rate (blue dots), with the black dots representing the original semi-major axis and eccentricity of asteroids in the present asteroid belt. 

Figure~\ref{tid_v_total_earth} shows the total number of particles in the $N$--body simulation and the number of tidal disruption events of the two simulations, with and without Earth engulfment after a time of $50\, \rm Myr$, all as a function of semi-major axis. The difference in the location of the peak represents the shift in location of the $\nu_6$ secular resonance caused by the engulfment of the Earth. The shift in the peak is approximately $0.05\,\rm au$, similar to the shift shown by the analytic model in Fig.~\ref{width}. This demonstrates that the engulfment of the Earth indeed shifts the resonance into a more highly populated region of the asteroid belt.

Figure~\ref{a_e_time} shows the initial distribution in semi--major axis and eccentricity of the objects that were tidally disrupted during the simulations  without Earth engulfment (left panel) and with Earth engulfment (right panel). There is a higher concentration of tidally disrupted objects with an initial location within the stable region of the asteroid belt when the Earth is engulfed. This comparison demonstrates that the change is caused by the shift in the $\nu_6$ secular resonance. At later times, the objects that are tidally disrupted mostly come from the location of the $\nu_6$ resonance. This shows that their eccentricity growth is indeed due to secular effects, contrary to the highly unstable population represented by green points in Fig.~\ref{a_e_time}. Comparing the two panels it is clear that the location of the resonance has shifted with the engulfment of the Earth.



Finally, Fig.~\ref{rate_hist} shows the tidal disruption rate for asteroids as a function of time for the evolution of asteroids near the $\nu_6$ secular resonance. We include the rates for the simulations with and without Earth engulfed and also show the difference between these two rates.  Furthermore, we show a constant best-fitting line of the difference that represents the continuous supply of asteroids coming from the $\nu_6$ resonance. The number of test particles that undergo tidal disruption is normalized to the initial number of particles in our simulations. There is  a continuous rate of tidal disruptions throughout the simulation. White dwarf pollution is observed at $30\, \rm Myr$ to $600\, \rm Myr$ cooling ages, thus our simulation time of $50$ million years is longer than the observed lower limit of pollution cooling age.

\begin{figure*}
\includegraphics[width=5.65cm]{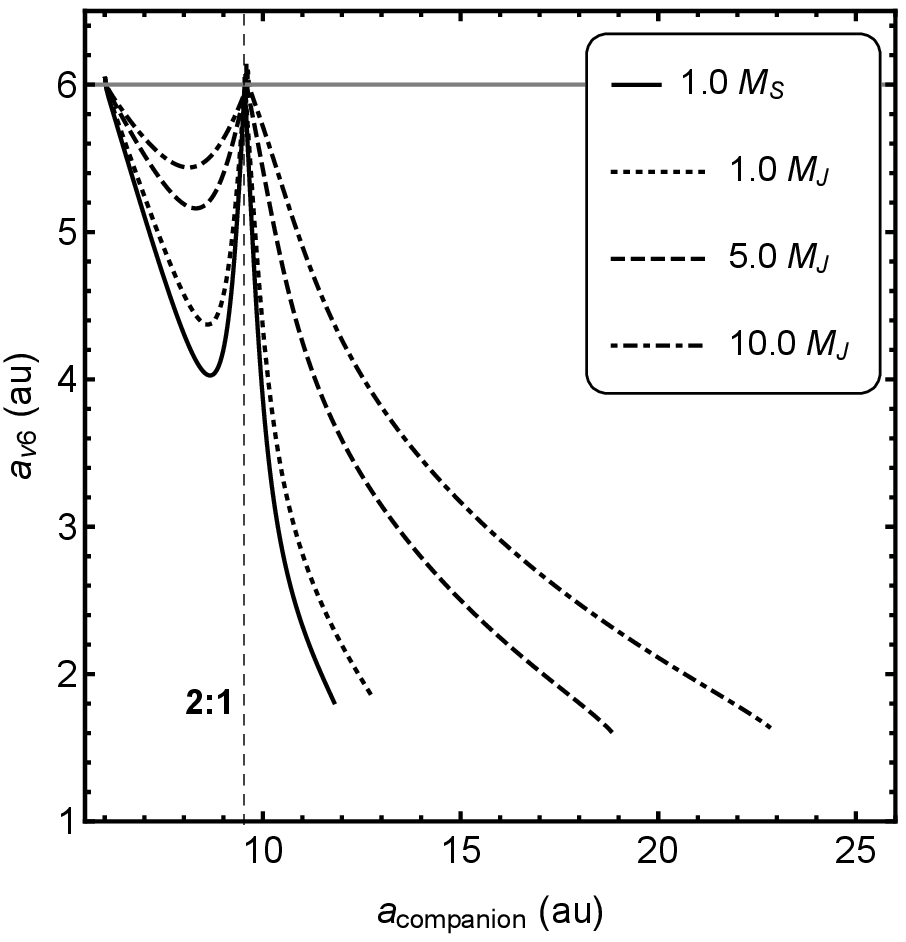}
\includegraphics[width=5.8cm]{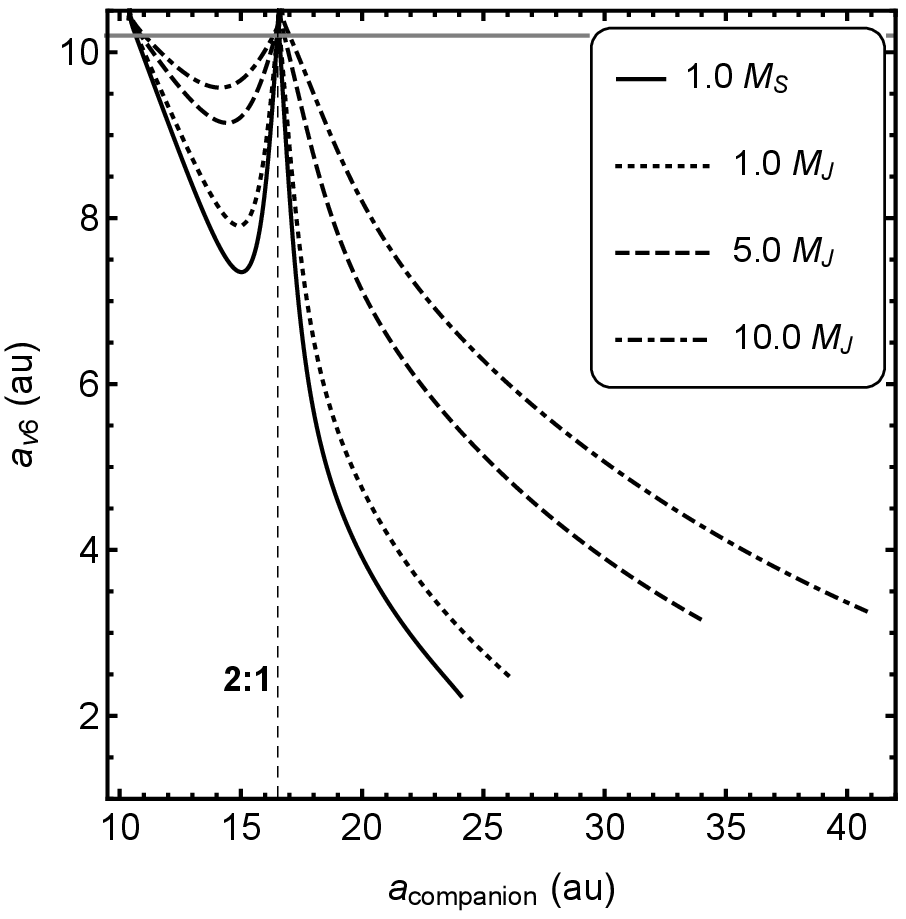}
\includegraphics[width=5.8cm]{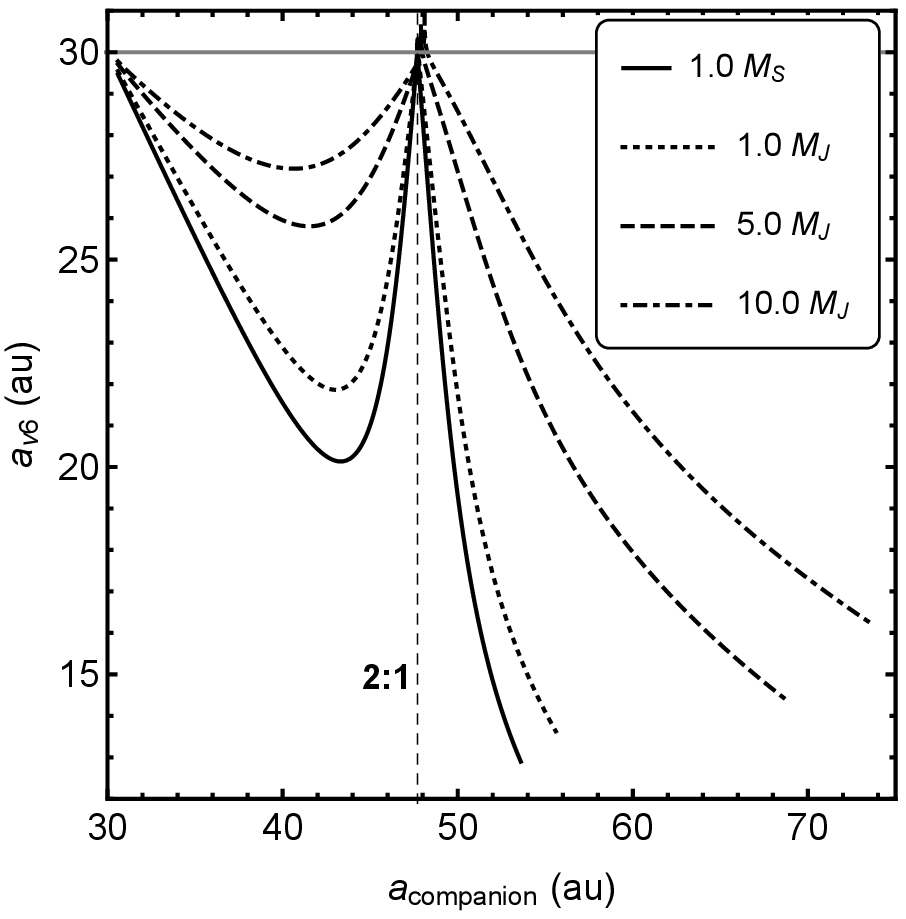}
\caption{Location of the $\nu_6$ secular resonance as a function of the semi-major axis of various planetary companions for white dwarf mass $M_{\rm wd} = 0.5\, \rm M_{\odot}$. The semi-major axis of the inner companion is constant at $6\, \rm  au$ (left panel), $10.4\, \rm  au$ (middle panel), and $30\, \rm  au$ (right panel). The masses of the outer planetary companions that are considered include $1.0$ Saturn mass ($M_{\rm s}$, solid), $1.0$ Jupiter mass ($M_{\rm J}$, dotted), $5.0\, M_{\rm J}$ (dashed), and $10.0\, M_{\rm J}$.  A correction was implemented due to the near 2:1 mean-motion resonance between Jupiter and Saturn \citep{Malhotra1989,Minton2011}. The vertical black-dotted line shows the location of this 2:1 mean-motion resonance and the semi-major axis of the inner Jupiter mass planetary companion is shown by the horizontal line. Note that relevant results do not hold for small semi-major axes of the outer planetary companion due to our first order approximation.}
\label{Saturn}
\end{figure*}

\begin{figure*}
\includegraphics[width=5.65cm]{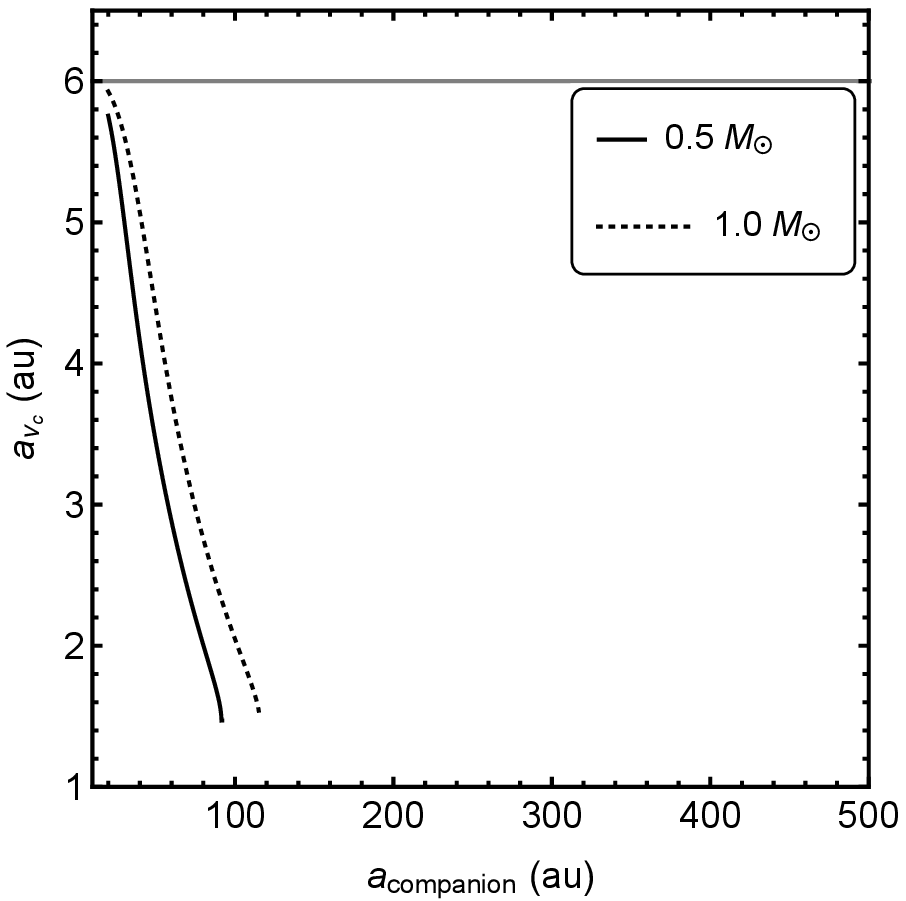}
\includegraphics[width=5.8cm]{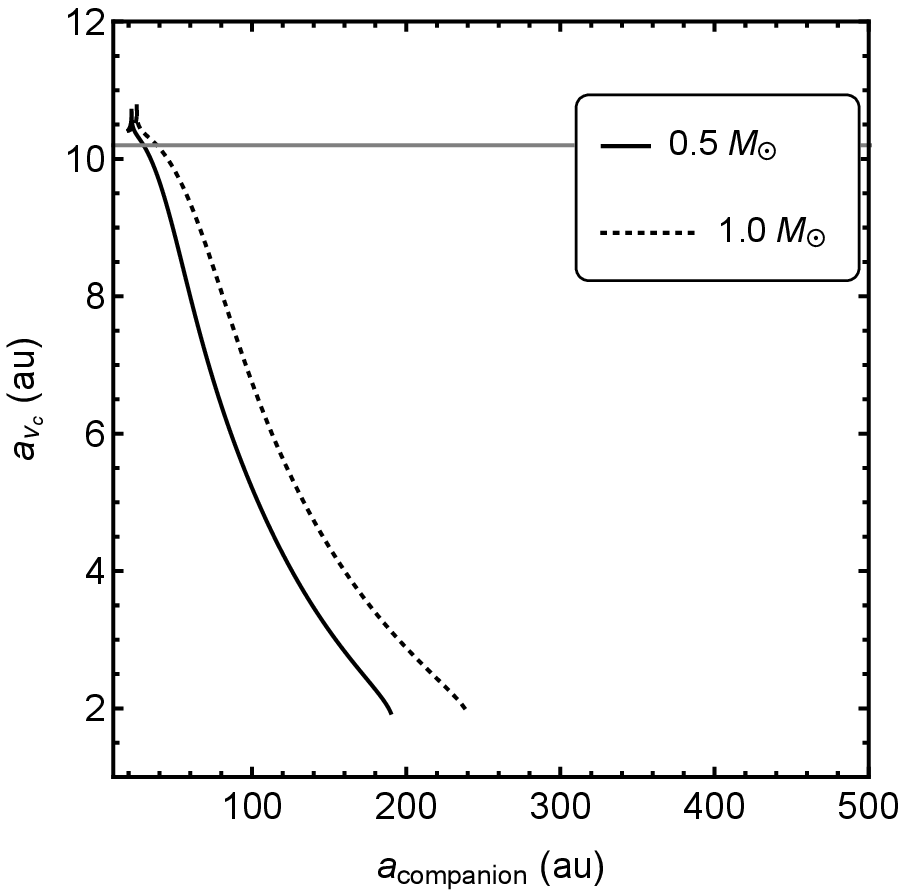}
\includegraphics[width=5.8cm]{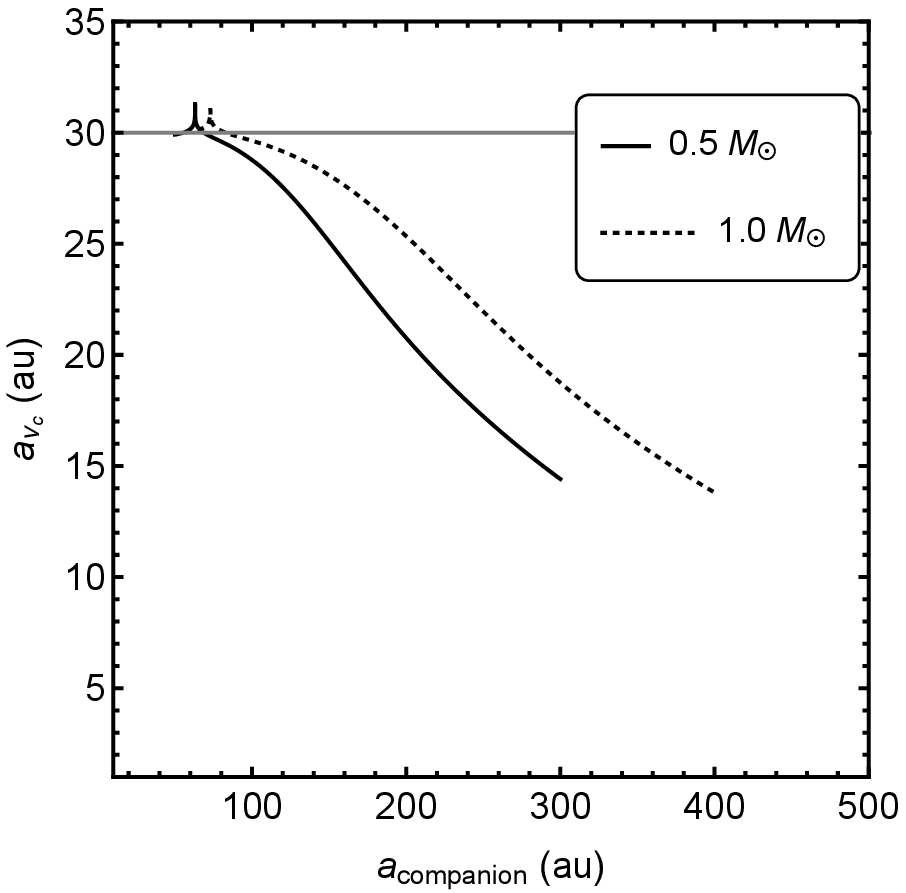}
\caption{The location of the companion ($\nu_c$) secular resonance as a function of the orbital separation of the stellar companion for various stellar companion masses. We also vary the semi-major axis of the inner planetary companion as follows: $6 \, \rm  au$ (left panel), $10.4 \, \rm au$ (middle panel), and $30 \, \rm  au$ (right panel).  The masses that were modeled include  $0.5 \, \rm M_{\odot}$ (solid) and $1.0 \, \rm M_{\odot}$ (dotted). The companion star is orbiting a $0.5\, \rm M_{\odot}$ white dwarf with a Jupiter-mass planet with a semi-major axis shown by the horizontal gray line. Note that relevant results do not hold for small semi-major axes of the stellar companion due to our first order approximation.
}
\label{binary}
\end{figure*}

We estimate the mass accretion rate for our secular resonance model based on a tidal disruption rate of  ${\cal{R}_{\rm D}} = 0.0002\, \rm Myr^{-1}$ per particle calculated from the  constant best-fitting line in Fig.~\ref{rate_hist} to be
\begin{align}
\dot{M}_{\rm acc} \simeq  & 1.0\times 10^{5} {\rm g\,s^{-1}}\bigg(\frac{N_{\rm tot}}{10000}\bigg) \bigg(\frac{\rho_{\rm ast}}{3{\rm g/cm^3}}\bigg) \bigg(\frac{R_{\rm ast}}{5 {\rm km}}\bigg)^{3} \notag \\
& \times \bigg(\frac{\mathcal{R_{\rm D}}}{0.0002\,{\rm Myr^{-1}}}\bigg),
\end{align}
where $N_{\rm tot}$ is the total number of asteroids near the $\nu_6$ secular resonance, $\rho_{\rm ast}$ is the average asteroid density, $R_{\rm ast}$ is the average asteroid radius. We assume an average density and radius of $3\,\rm g/cm^3$ and $5\, \rm km$, respectively \cite[e.g.][]{Reach2005}. The estimate for the mass accretion rate in our model is within the  range of the observed accretion rates calculated by \cite{Koester2014} \cite[see also Figure 10 in][]{Farihi2016}. This estimated accretion rate should be considered as a lower limit because we have only simulated a small portion of the asteroid belt. Other resonances may also play a (possibly small) role, for example the 2:1 resonance \citep{Debes2012}. Furthermore, the asteroid belt in the Solar system is much less massive than other known debris discs. When the planetary debris particles enter this tidal disruption zone  they will be torn apart, forming a debris disc with asteroidal composition, and eventually be accreted on the white dwarf. This debris disc will act as a reservoir which could produce a steadier supply of heavy elements at long timescales \citep{Deal2013}. Thus the rates shown in Fig.~\ref{rate_hist} should be lower, but probably steadier.


The results of our 4--body simulations agree with the analytic model presented in Section~{\ref{sec:analytic}},  to the extent that we observe the shift of the secular resonance. The observed difference in the number of tidal disruption events in Fig.~\ref{tid_v_total_earth} and the shift in the concentration of tidal disruption events in Fig.~\ref{a_e_time} show that the location of the $\nu_6$ resonance has shifted outwards by about $0.05\, \rm au$ into the asteroid belt as predicted analytically in Figure~\ref{width}. In the next section we consider how secular perturbations may apply more generally to exoplanetary systems with the analytic model of the secular resonance.

\section{Exoplanetary Systems}
\label{sec:exo}
Secular resonances are sensitive to the architecture of a planetary system \citep[e.g][]{Minton2011,smallwood2017}. In this Section we consider how the $\nu_6$ secular resonance may pollute a white dwarf for different planetary architectures with the analytic model described in  Section~\ref{sec:analytic}.
 First, we look at the displacement of the $\nu_6$ secular resonance for varying mass and location of Saturn in the Solar System. Next, we examine the location of secular resonances in planetary systems with a binary star companion. Each model does not include an inner Earth-like terrestrial planet because the secular resonance shift in amplitude is small. We focus solely  on the location of the secular resonance which is important for increasing eccentricities of asteroids.



\subsection{Planetary Companions}
\label{sec:planetary}
Here we examine a system with two outer giant planetary companions around a white dwarf. In order to generalize our results to exoplanetary systems, we  calculated how  the resonance location changes with the semi-major axis and mass of the outer planetary companion.  We model three architectures with the inner planetary companion being kept as a Jupiter-mass planet with semi-major axes $6\, \rm au$,  $10.4\, \rm au$, and $30\, \rm au$.  The semi-major axes larger than and smaller than Jupiter's adiabatic semi-major axis are taken as fiducial estimates to test the dynamics of secular resonances in varying planetary architectures.

The location of the secular resonance as a function of the outer companion's semi-major axis was found by calculating the resulting eigenfrequency 
and then finding the location of the intersection with the free precession rate of a test particle.  
We included a correction due to the near 2:1 mean-motion resonance between the two companions \citep{Malhotra1989,Minton2011}. 

Figure~\ref{Saturn} shows the location of the $\nu_6$ secular resonance for three different architectures as a function of  planetary companion semi-major axis with a variety of planetary masses for the outer companion that include $1.0$ Saturn mass ($\rm M_{S}$), $1.0$ Jupiter mass ($\rm M_{J}$), $5.0 \, \rm M_{J}$, and  $10.0 \, \rm M_{J}$.  We consider the case where the planetary companion is orbiting a $0.5 \, \rm M_{\odot}$ white dwarf. The inner and outer companions orbits expand adiabatically in response to the amount of stellar mass loss. For comparison, the inner planetary companion's semi-major axis is denoted in Fig.~\ref{Saturn} by the horizontal gray line. Let us consider an asteroid belt initially confined by the $\nu_6$-like secular resonance. Assuming that an efficient rate of disrupted asteroids is obtained whenever the asteroid belt is close enough to the star, the more massive the outer planetary companion, the larger its semi-major axis.

The analytical models suggest that for the case of at least two surviving giant planets orbiting a white dwarf, the $\nu_6$ secular resonance can exist among a variety of exoplanetary architecture. The outward shift of secular resonances within exoplanetary systems would arise from the engulfment of terrestrial planets located near the host star, leading to the formation of debris discs around white dwarf stars and subsequent pollution of their atmosphere.

\subsection{Stellar Companions}
\label{sec:stellar}

Roughly $50\%$ of stars in the Milky Way are in binary systems \citep{Horch2014}.  Many polluted white dwarfs are also observed in  binary systems \citep{Zuckerman2003}. The proposed theoretical models for white dwarf pollution in binaries include  perturbations by  galactic tides for wide binaries \citep{Bonsor2015} and  Kozai-Lidov oscillations  \citep{Hamers2016,Petrovich2017}. Here we consider closer binaries that are close to coplanar to the planetary system for which none of these mechanisms are possible.

To identify how the secular resonance operates in a binary system, we  use our analytic model described in Sec.~\ref{sec:analytic}. We replace the outer planetary companion with a stellar companion. In Fig.~\ref{binary} we vary the mass and semi-major axis of the companion star for a $0.5 \, \rm M_{\odot}$ white dwarf and calculate the location of the resulting secular resonance for three different semi-major axis values of the inner Jupiter mass planet.  In Fig.~\ref{binary}, the location of the inner planet is $6.0\, \rm  au$ (left panel),  $10.4\, \rm  au$ (middle panel), and $30\, \rm  au$ (right panel).  
In each case, we vary the mass of the companion star as listed:  $0.5\, \rm M_{\odot}$ (solid line) and $1.0\, \rm M_{\odot}$ (dashed). The middle panel corresponds to Jupiter at $5.2\, \rm au$ initially. In each panel, as the mass of the stellar companion increases, the location of the secular resonance can exist at a wider binary separations. 
Assuming that an asteroid belt is initially confined by the $\nu_6$-like secular resonance, these models demonstrate that a variety of binary configurations may produce white dwarf pollution. 
Depending on the semi-major axis of the giant planet, this pollution mechanism (of secular perturbations) can support white dwarf pollution located in binaries with a binary separation $< 400\,\rm au$. Note that since the Laplace-Lagrange equations are first order with respect to the orbital perturbations, this holds true only if this massive companion is very far way. We speculate that a white dwarf within a wide binary (i.e. $a_{\rm binary} > 400\,\rm  au$) can still become polluted not by the binary companion itself, but by perturbations driven by surviving planets orbiting the white dwarf, which follows the processes described in Section~\ref{sec:analytic}. 




\section{Conclusions}


We have found that secular resonances, in particular the $\nu_6$ resonance, can be responsible for the presence of heavy elements in the atmosphere of white dwarfs.
In the solar system, when the Earth becomes engulfed by the Sun as the latter leaves the main--sequence, the $\nu_6$ resonance is shifted outward. This mainly occurs because the free precession frequency of the asteroid is changed. This change in the location of the resonance causes previously stable asteroids to undergo secular resonant perturbations that lead to a higher rate of tidal disruptions close to the white dwarf. The resulting debris disc of heavy elements accretes onto the white dwarf, polluting the atmosphere. This mechanism can lead to white dwarf pollution for a large range of planetary system parameters including systems with two giant planets, or one planet and a binary star companion, assuming the presence of an asteroid belt initially confined by secular resonances, and the engulfment of an inner planet during the evolution of the star.
From our numerical experiments, we expect the process of asteroid perturbations by secular resonances to last much longer than the white dwarf cooling age given a massive enough asteroid belt.

\cite{Debes2012} modeled the change in width of the 2:1 mean-motion resonance and found that the current mass of the asteroid belt would need to increase by a factor of $10^2 - 10^4$ in order for sufficient material to be accreted by the white dwarf by this mechanism alone.  \cite{Frewen2014} later found that their model, that dealt with the accretion of a single planet, required a planetesimal disc to be a few thousand times larger than the asteroid belt. The estimate for the mass accretion rate in our model is on the lower end of the range of the observed accretion rates calculated by \cite{Koester2014} \cite[see also Figure 10 in][]{Farihi2016}. However, the asteroid belt in our Solar system is much less massive than other known warm debris discs. Knowing that there exist more massive planetesimal belts than our asteroid belt \citep{Moro-mart2010} gives encouraging evidence that secular resonances can potentially pollute white dwarfs.
   
While we don't know with certainty whether the Sun will be polluted during its white dwarf stage due to asteroidal accretion, the analytical and numerical models used in this work  do provide a possible mechanism for white dwarf pollution in exoplanetary systems.  There is probably not just one mechanism that produces white dwarf pollution. Theoretically, our secular resonance model and the \cite{Debes2012} mean-motion resonance model could operate in a synergistic manner,  allowing a larger fraction of asteroids to become tidally disrupted,  but for exoplanetary systems resembling the Solar system, we expect that a higher flux of asteroids is influenced by secular perturbations rather than by mean-motion perturbations.

There are many known planetary system architectures in various databases including Kepler \citep{Borucki2010,Borucki2011,Batalha2013},  CoRoT \citep{Auvergne2009}, SuperWASP \citep{Pollacco2006}, and KELT \citep{Pepper2007}. In our pollution model we focus on exoplanetary systems that share strong features with the Solar system, including: two outer massive planets, an inner asteroid belt truncated by a secular resonance, an inner planet that is engulfed during the stellar evolution, and where all the bodies are on nearly circular and coplanar orbits. White dwarf pollution occurs in planetary systems that are vastly different from our Solar System. However, secular and mean-motion resonances are expected to sculpt the architecture of any asteroid belt in exoplanetary systems, so the global mechanisms presented in this article should be triggered (with more or less efficiency) in various planetary configurations.



\section*{Acknowledgements}
We thank the anonymous referee for the helpful feedback which improved the quality of the manuscript. JLS also acknowledges support from a graduate fellowship from the Nevada Space Grant Consortium (NVSGC). All the simulations were submitted to the UNLV National Supercomputing Institute on the Cherry Creek cluster.  We acknowledge support from NASA through grant NNX17AB96G.











\bibliographystyle{mnras}
\bibliography{smallwood}

\bsp	
\label{lastpage}
\end{document}